# High Performance CNFET-based Ternary Full Adders

Fazel Sharifi, Atiyeh Panahi, Mohammad Hossein Moaiyeri and Keivan Navi

Nano-technology & Quantum Computing lab, Shahid Beheshti University, GC, Tehran, Iran. e-mails: f_sharifi@sbu.ac.ir, at.panahi@mail.sbu.ac.ir, h_moaiyeri@sbu.ac.ir and navi@sbu.ac.ir.

## ABSTRACT

This paper investigates the use of carbon nanotube field effect transistors (CNFETs) for the design of ternary full adder cells. The proposed circuits have been designed based on the unique properties of CNFETs such as having desired threshold voltages by adjusting diameter of the CNFETs gate nanotubes. The proposed circuits are examined using HSPICE simulator with the standard 32 nm CNFET technology. The proposed methods are simulated at different conditions such as different supply voltages, different temperature and operational frequencies. Simulation results show that the proposed designs are faster than the state of the art CNFET based ternary full adders.

.

## 1. INTRODUCTION

Due to the limitations of CMOS transistors, they are not able to continue the process of feature size reduction and should be replaced by new alternative emerging technologies. CMOS transistors have problems such as short channel effect, reduced gate control, high leakage power and parameter variation [1]. So thinking about alternative new technologies such as Quantum dot Cellular Automata (QCA) [2], Single Electron Transistors (SET) [3] and Carbon Nanotube Field-Effect Transistors (CNFET) [4] is essential.

A suitable alternative for CMOS transistors is CNFET. Because of the similarities between CMOS and CNFET transistors in terms of inherent electronic parameters, CNFET could be a good alternative technology without any major changes in CMOS platforms [4]. In addition, a unique characteristic of CNFET devices is one dimensional band structure which suppresses backscattering and causes near ballistic operation, that makes it suitable for implementing fast and low power CNFET based circuits [4]. Another feature of CNFET is that it has same mobility and consequently same current drive for P-FET and N-FET devices. This makes transistor sizing easier for complex circuits [5]. Among all applications of CNFET transistors, Multiple Valued Logic (MVL) could be more of an interest. MVL logic means using more than two logic values for designing circuits and systems.

Using CNFETs is appropriate for MVL designing. Because MVL is based on multiple threshold design technique and determining the threshold voltage of CNFETs is easily possible by changing the diameter of the nanotubes [10]. One of the major challenges of binary logic is the number of pin counts and interconnects specially in dense chips. This problem limits the number of inside and outside connections [7]. By using MVL, we can reduce the circuit area by reducing the overhead in interconnects and pin counts [7]. In MVL designs, wires and interconnections carry more information than binary logic; so it has higher speed and smaller number of computation stages [6]. Among all radices that exist for MVL logic, e (≈ 2.718) base operations have the most efficient implementation [8]. But due to the hardware restrictions for implementing real systems, we should use natural numbers as the base of computations. So radix 3 which is the nearest natural number to e; is more attractive and as a result, ternary logic is the best and leads to less complexity and production cost [9].

Since adders are one of the most basic functional units in computer arithmetic and using high speed and low power adders can improve efficiency of other operations, in this paper two new efficient CNFET based ternary full adders are proposed.

This paper is organized as follows: the next section describes the CNFET device characteristics in detail. Section 3 presents the proposed methods for ternary full adder cells and section 4 compares these methods with other existing ternary full adders and includes the simulation results and finally section 5 concludes the paper.

## 2. CARBON NANOTUBE FIELD EFFECT TRANSISTOR (CNFET)

Carbon Nano Tube Field Effect Transistors (CNFET) are formed in the shape of a sheet of graphite tubes. Some advantages of CNFETs are such as they have higher ON current compared to MOSFET transistors. By using CNFETs it is possible to scale down feature size, beyond what currently lithographic methods permit. Also ballistic conduction of CNFETs reduces the power dissipation in the transistor body. One dimension structure of CNTs reduces the resistivity and consequently the energy dissipation and power consumption [7].

CNTs are grouped to Single-Walled Carbon Nano Tube (SWCNT) and Multi-Walled Carbon Nano Tube (MWCNT). SWCNTs are made of one cylinder and MWCNTs are made of more than one cylinder that are nested inside each other [9]. SWCNT could be conductor or semiconductor. Electrical and physical properties of CNFETs are depending on the orientation of chiral vector components and the rolling direction of graphite sheet [11]. Chiral vector describes structure of CNTs and is determined by ($n_1$, $n_2$) indices. For example if $n_1$-$n_2$ = 3k (k∈Z), SWCNT has the metallic characteristics, otherwise it is semiconductor [4].

Several SWCNTs could be placed next to each other under the transistor gate and set its width. The width of a CNFET transistor depends on the number of tubes which are placed under the transistor gate and also it depends on the distance between two adjacent tubes which is called a pitch. Therefore the width of a transistor is determined by the following equation [12].

$$W_{gate} \cong Min(W_{min}, N \times Pitch) \quad (1)$$

Where, $N$ is the number of nanotubes that are placed under the transistor gate and $W_{min}$ is the minimum width of the gate.

Threshold voltage of CNFET transistors is determined by the following equation [12].

$$V_{th} \cong \frac{E_g}{2e} = \frac{\sqrt{3}}{3}\frac{aV_\pi}{eD_{CNT}} \cong \frac{0.43}{D_{CNT}(nm)} \quad (2)$$

In the above mentioned equation, $a$ is the carbon to carbon atom distance, $V_\pi$ is the carbon $\pi - \pi$ band energy in the tight bonding model, $e$ is the unit electron charge and $D_{CNT}$ is the diameter of CNFETs that can be calculated by the following equation [12].

$$D_{CNT} = \frac{a \times \sqrt{{n_1}^2 + {n_2}^2 + n_1 n_2}}{\pi} \cong 0.0783\sqrt{{n_1}^2 + {n_2}^2 + n_1 n_2} \quad (3)$$

As previously mentioned and due to the Equation (2) and Equation (3), modification of CNFET threshold voltage is possible only by changing the diameter of the nanotubes. So CNFETs are appropriate for implementing multiple threshold circuits. By changing the chiral vector indices the nanotube diameter of transistor changes and consequently the threshold voltage of CNFET sets simply.

Three different types of CNFETs which have been introduced are as following. The first type is Schottky-Barrier (SB) that is created by tunneling of electrons through Schottky-Barrier at the source-channel junction. The source and drain of this type of CNFETs are metallic inside a semiconductor body; the direct contact between metal and semiconductor causes strong ambipolar characteristic which restricts the usage of this type in CMOS-like logic families. The second type of CNFETs is called MOSFET-like. The source and drain of this type are heavily doped and they have field effect and unipolar characteristics. In general, this type of CNFETs operates like the normal MOSFETs but its performance is better. Because of the high ON current in this type of CNFETs, they are appropriate for ultra-high performance circuits. Band-to-band CNFET transistors are the third type of CNFETs (T-CNFET). Due to super cut-off characteristic of them, they have been used for sub-threshold and ultra-low-power applications [13].

In addition to all the benefits of CNFETs, fabrication problems such as placing CNFETs on an existing MOSFET platform is one of the challenges in using them [14].

Due to the similarities between MOSFET-like CNTs and MOSFETs in terms of inherent characteristics, MOSFET-like CNTs are utilized in this paper for designing the proposed full adder cells.

## 3. PROPOSED WORK

Ternary logic that is the most important type of MVLs, is made of three logic levels that represent themselves with "0", "1" and "2" logical values. Equivalent voltage of these logical values are "0", "$\frac{V_{dd}}{2}$" and "$V_{dd}$". As previously mentioned, the full adder cell is one of the most important functional units of arithmetic operations. So in this section two proposed designs for the ternary full adder cell are presented. In a ternary full adder the relation between inputs and outputs represents by the following equation:

$$\sum in = A + B + C_{in} = 3C_{out} + Sum \quad (4)$$

Therefor:

$$Sum = \sum in - 3C_{out} \quad (5)$$

Where, $A$ and $B$ are input trits (TRinary digITal unit), $C_{in}$ is the input carry and $C_{out}$ is the output carry trit that weighs 3. In the ternary logic, operations are done at radix 3 and consequently weight of each trit position is 3 times more than the weight of previous trit position.

According to Equation (4) and also by considering Equation (6) for a given dividend as $X$, a divisor as $D$, a quotient as $Q$ and a reminder as $R$ [17],

$$X = QD + R \;\; with\, R < D \quad (6)$$

if we suppose that $X = A + B + C_{in} = \sum in$ and $D = 3$, it is concluded that the Sum output trit is the reminder of the division and $C_{out}$ trit is the quotient of the division. So:

$$\begin{cases} Sum = \sum in \bmod 3 \\ C_{out} = \left\lfloor \frac{\sum in}{3} \right\rfloor \end{cases} \quad (7)$$

At the above equation, ⌊ ⌋ notation represents the floor function that results the integer part of the quotient of division. As a result:

$$C_{out} = \begin{cases} 0 & 0 \le \sum in \le 2 \\ 1 & 3 \le \sum in \le 5 \\ 2 & 5 < \sum in \le 6 \end{cases} \quad (8)$$

According to Equation (5) and Equation (8) for the Sum output trit the following equation is concluded:

$$Sum = \sum in - 3C_{out} = \begin{cases} \sum in & 0 \le \sum in \le 2 \\ \sum in - 3 & 3 \le \sum in \le 5 \\ \sum in - 6 & 5 < \sum in \le 6 \end{cases} \quad (9)$$

Truth table of the ternary full adder is represented in Table 1. According to the Table 1 and Equation (8), $C_{out}$ trit could be implemented with a ternary buffer. Threshold voltages of $C_{out}$ buffer are determined easily by setting nanotube diameters of CNFET transistors. This ternary buffer represented in Figure 1. Its threshold voltages have been set so that to implement the Equation (8). Logical values of the inputs of this buffer would be one of the numbers that is shown in Figure 1. If the logical values of the input trits ($\sum in$) be less than 2.5 then the output will be equal to "0" and similarity, if $\sum in > 5.5$, upper path of output will be active and output node would be equal to "$V_{dd}$". For other values of $\sum in$, two paths would be active and logical output value would be "1" and voltage value of this node would be equal to "$\frac{V_{dd}}{2}$".

Table 1: Truth table of ternary full adder

| $A + B + C_{in}$ | $Sum$ | $C_{out}$ |
|---|---|---|
| 0 | 0 | 0 |
| 1 | 1 | 0 |
| 2 | 2 | 0 |
| 3 | 0 | 1 |
| 4 | 1 | 1 |
| 5 | 2 | 1 |
| 6 | 0 | 2 |

According to Table 1 and Equation (9), the Sum output determines by the output carry trit and summation of the logical values of the input trits ($\sum in$). The proposed methods are based on this idea that the $C_{out}$ trit determines that for each of the input trits, which paths of Sum output should be activated. The first proposed design of the Sum output is shown in Figure 2. In this figure $S$ and $F$ signals are selectors that determine output path. $S$ is the output of a binary inverter with threshold logic value of 2.5 and also $F$ signal is the output of a binary inverter with threshold logic value of 5.5. So the Sum output trit is determined based on the $C_{out}$ trit and the values of these control signals. They specify that for each combination of the input trits which of the paths should be activated.

According to the Figure 2 for the first proposed Sum output, If $\sum in < 2.5$ so based on the Equation (8), $C_{out}$= 0 and $S$ = 1. Therefore $TG_0$ (Transmission Gate) is ON and the first path will be active and output of the $STI_0$ gate after passing through a ternary inverter reaches to output. $STI_0$ is a Standard Ternary Inverter that operates on "0", "1" and "2" logic values. Threshold logic values of PFET and NFET transistors in this standard ternary inverter are set so that to implement this functionality. If $2.5 < \sum in < 5.5$ then according to Equation (8), $C_{out}$ = 1, $S$ = 0 and $F$ = 1 so $TG_1$ and $TG_2$ are active and output of $STI_1$ that operates on logic values of "3", "4" and "5", reaches to output. The output of $STI_1$ gate is logically equal to "0", "1" and "2" for input logic values of "3", "4" and "5". And finally, if $5.5 < \sum in$, so $F = 0$, $\bar{F} = 1$ and NFET transistor is active and Sum output becomes "0".
Therefore the Sum output trit determines according to $C_{out}$ and also consequently $S$ and $F$ selector signals. In another word, for each of the input combinations, one of the paths will be activated and consequently there is no contention for the Sum output voltage. To reduce the number of transistors that are used in the first proposed design, it is possible to use a ternary buffer instead of two cascaded inverters. Schematic gate level design of the second proposed method is shown in Figure 3. The second proposed method has the same operation of the first proposed method as described previously. The only difference between two designs is that two cascaded inverter gates are replaced with $STB_0$and $STB_1$ternary buffer gates. To compare the time and area complexity of two proposed designs with each other, the first design is composed of 55 CNFETs and 3 input capacitors and the second design has 43 CNFETs and 3 input capacitors. So the second design has smaller area and also it has less delay. Because its Sum output transmits through one buffer gate whereas in the first design, the Sum output signal passes through two cascaded inverter gates which increases the critical path delay.

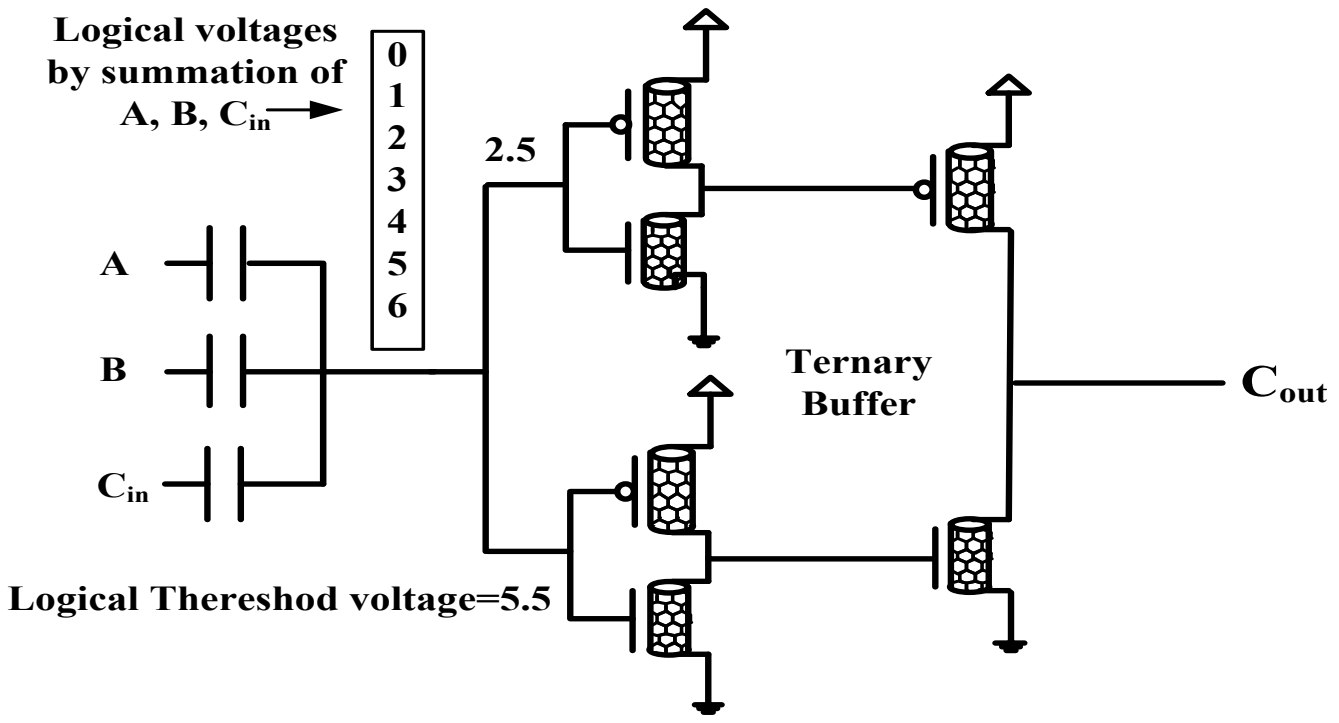

Figure 1: Output carry design.

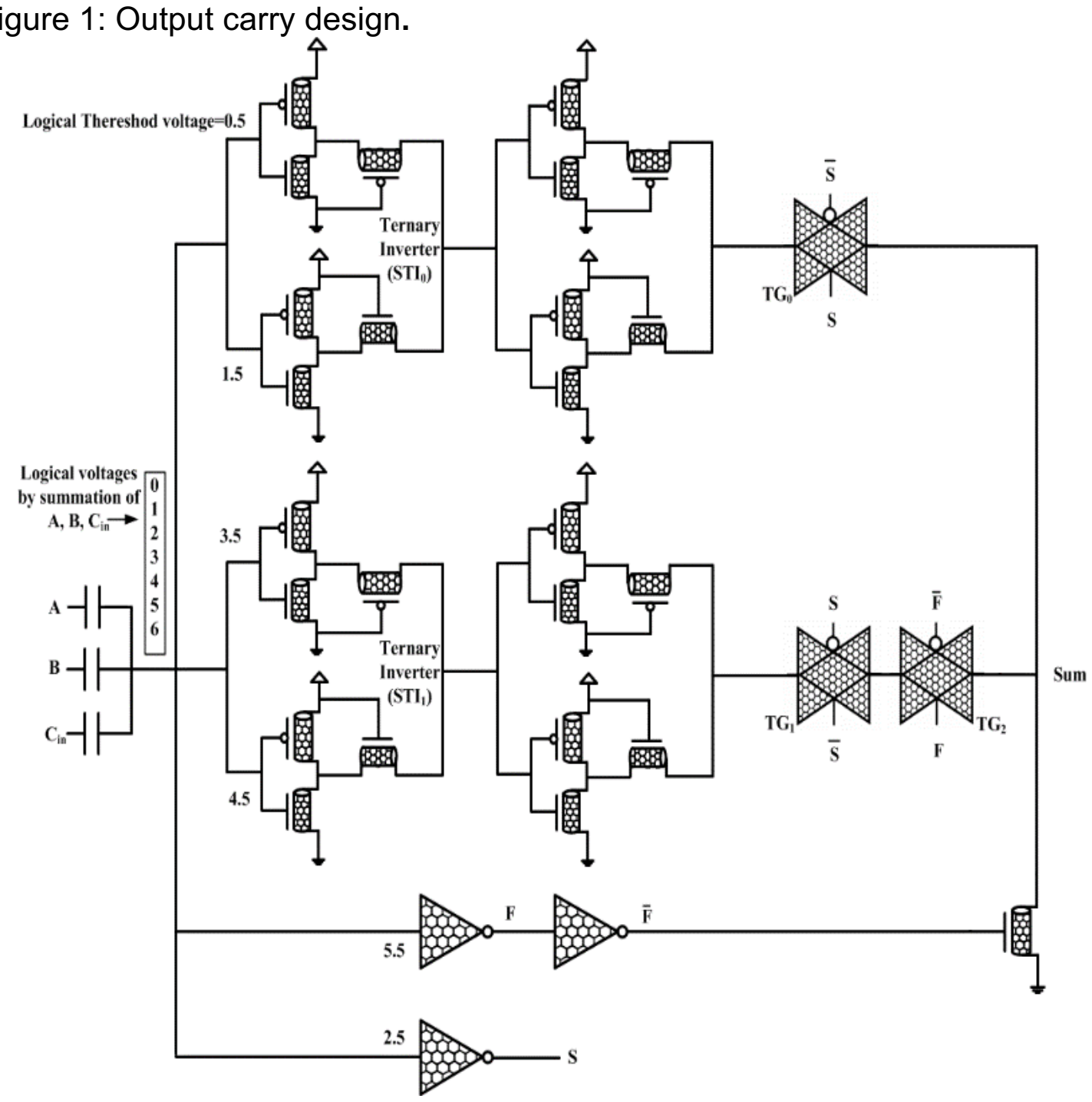

Figure 2: First proposed ternary full adder.

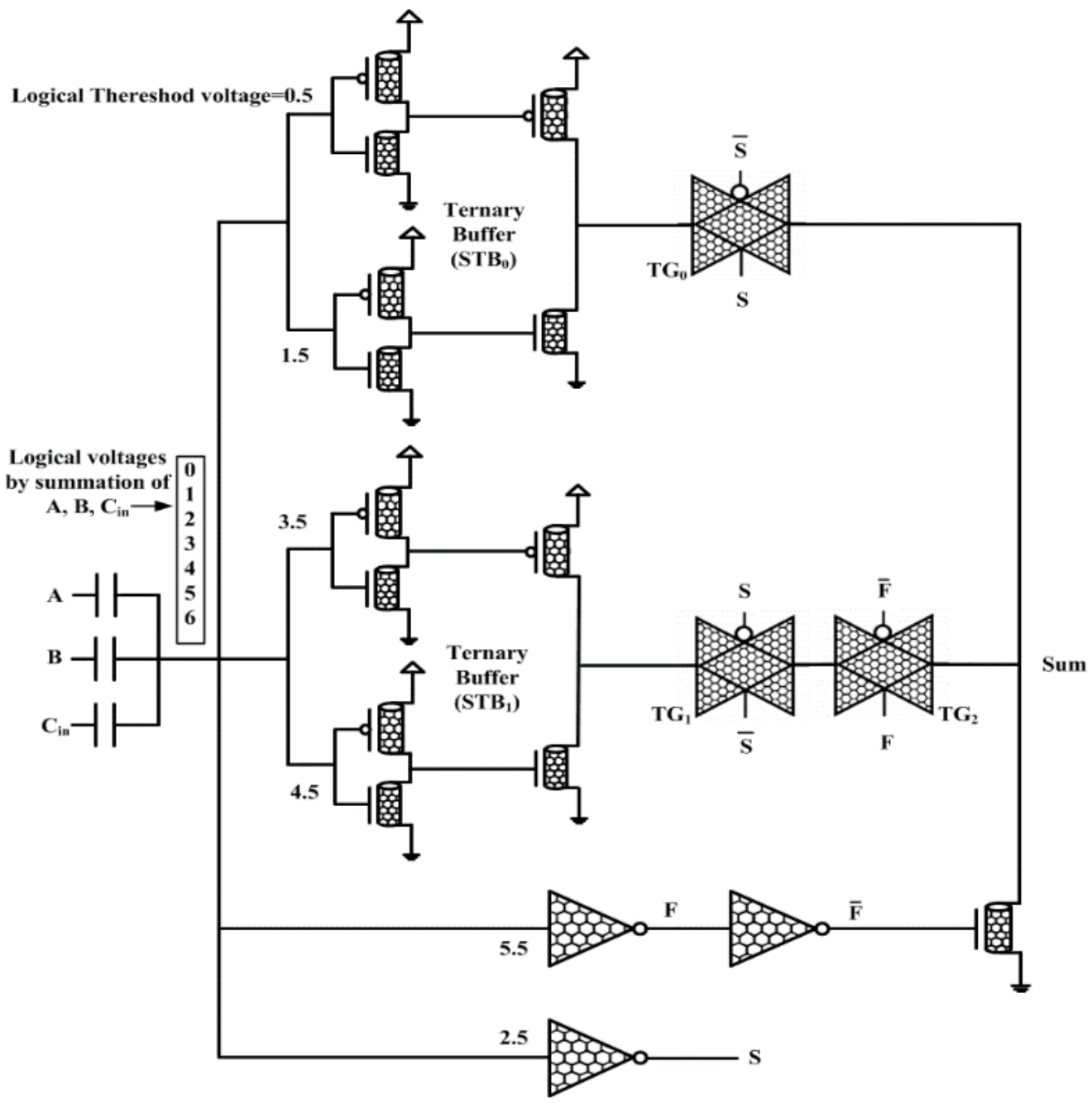

Figure 3: Second proposed ternary full adder. Table 2: CNFET model parameters [18]

## 4. Simulation Results

In this section, the results of simulating two proposed methods are presented. Simulation is done using Synopsys HSPICE simulator for 32 nm CMOS technology with Compact SPICE model for CNFET that is further described in [15] and [16]. Table 2 represents some of the important parameters of this CNFET model. The proposed methods are simulated at different conditions such as different supply voltages, different temperature and operational frequencies and also for different load capacitors in output nodes.
Figure 4 shows the waveform of output nodes for different combinations of inputs which confirms correct operating of ternary full adder.
In the following of this paper, we compared two proposed methods with other existing ternary full adder designs in different conditions. Average power consumption, worth case delay and power-delay product (PDP) are evaluation criteria of the circuits. Table 3 shows these parameters for the proposed methods versus methods presented in [7] at 100 MHz and 250 MHz operational frequencies and for a fixed 1fF load capacitor at output nodes. As Table 3 represents, two proposed methods have less delay and energy consumption comparing to proposed designs of [7]. For example at 250 MHz operational frequency, the first proposed method decreases the PDP up to 57.67% and the second proposed method reduces it up to 37.50%.
To compare the performance of the designs, ternary full adders are simulated at 250 MHz operational frequencies by three evaluation criteria (power, delay and PDP). To evaluate the operation of methods versus different load capacitors, load capacitors up to 5 fF are placed at the output nodes of the circuits. Figure 5 shows the simulation results. As it is clear, the proposed methods reduce the delay and energy consumption in the presence of different load capacitors. Figure 6 evaluates operation of the proposed methods by increasing the temperature at 500 MHz operational frequency. As anticipated, the operation of CNFETs is independent of the temperature changes. As Figure 6 shows, for different temperatures that are ranging from 0 ℃ to 80 ℃, all the three evaluated parameters are fixed without any major changes versus temperature. To evaluate the operation of methods versus different supply voltages, they are simulated at different supply voltages at 250 MHz operational frequencies for a fixed 1 fF load at output nodes. The results are represented in Figure 7. As shown in Figure 7, for all of the supply voltages the proposed methods have less delay and also less energy consumption comparing to proposed methods of [7].

Table 1: Truth table of ternary full adder

| Parameter | Description | Value |
|---|---|---|
| $L_{ch}$ | Physical channel Length | 32 nm |
| $L_{geff}$ | The mean free path in the intrinsic CNT channel | 100 nm |
| $L_{dd}$ | The length of doped CNT drain-side extension region | 32 nm |
| $L_{ss}$ | The length of doped CNT source-side extension region | 32 nm |
| $T_{ox}$ | The thickness of high-k top gate dielectric material | 1 nm |
| $K_{gate}$ | The dielectric constant of high-k top gate dielectric material | 16 |
| $E_{fi}$ | The Femi level's doped S/D tube | 6 eV |
| $C_{sub}$ | The coupling capacitor between the channel region and the substrate | 20 pF/m |

Table 3: simulation results of the methods

| Frequency = 250 MHz, Load = 1 fF, $V_{dd}$= 0.9 V | | | |
|---|---|---|---|
| | Delay (e-10 s) | Power ($e-6$ w) | PDP ($e-15$ J) |
| 1st Proposed | 0.665 | 10.412 | 0.692 |
| 2nd Proposed | 0.555 | 18.414 | 1.023 |
| 1st design of [7] | 2.531 | 6.4661 | 1.636 |
| 2nd design of [7] | 1.749 | 17.184 | 3.005 |
| Frequency = 100 MHz, Load =1 fF, $V_{dd}$= 0.9 V | | | |
| 1st Proposed | 0.665 | 10.16 | 0.676 |
| 2nd Proposed | 0.559 | 19.36 | 1.084 |
| 1st design of [7] | 2.599 | 6.380 | 1.658 |
| 2nd design of [7] | 1.755 | 17.35 | 3.046 |

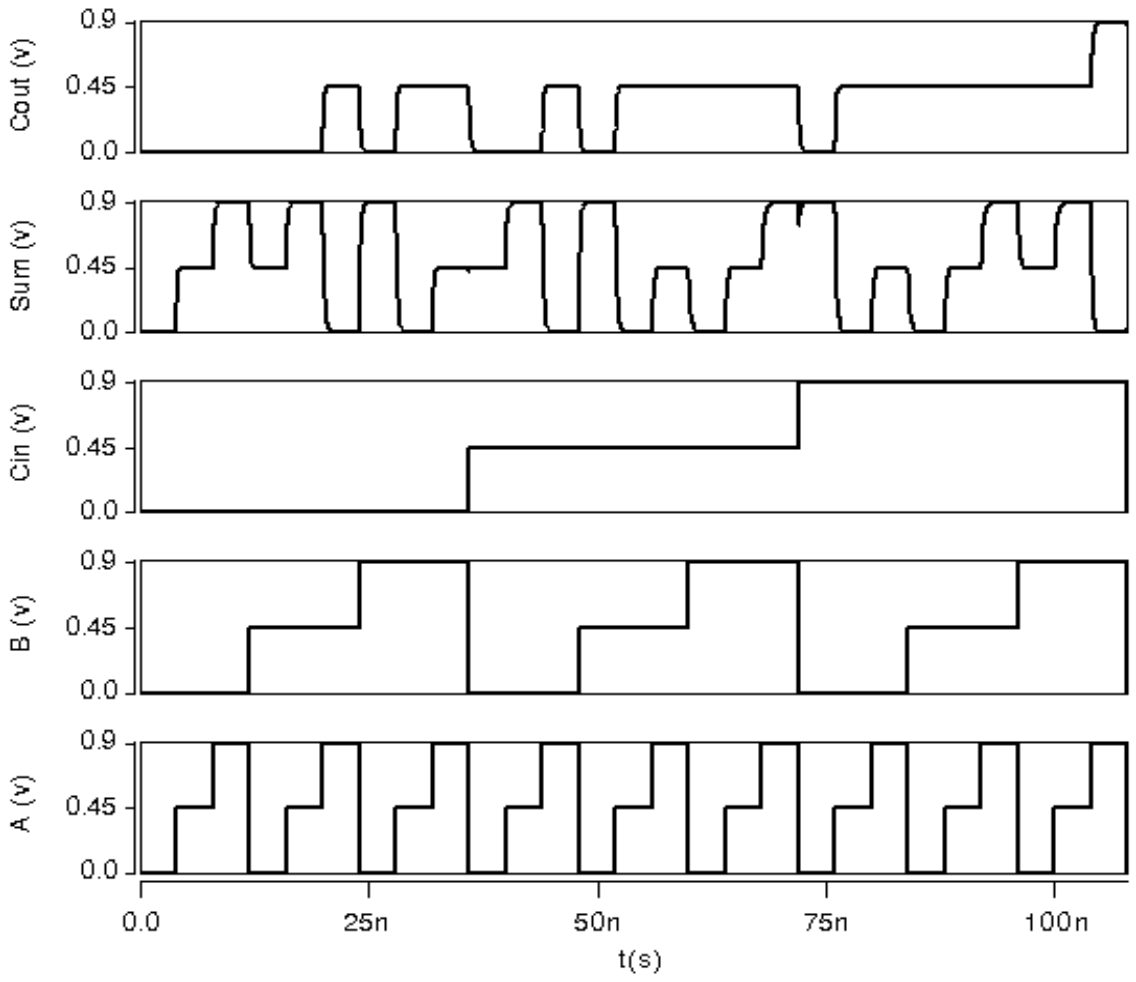

Figure 4: Transient response of the proposed design.

Average power consumption of the proposed methods is between the power consumption of the methods presented in [7]. It should be noted that the methods represented in [7] do not work properly for a supply voltage of 0.8 V. Therefore there is no data for these methods at this supply voltage.

## 6. CONCLUSION

Two novel high-performance ternary full adder cells have been proposed based on CNFETs. The proposed circuits have been designed based on multiple-Vth devices by utilizing unique characteristics of CNFETs. Simulation results indicate the superiority of the proposed designs in terms of delay and PDP compared to the other existing circuits in various conditions.

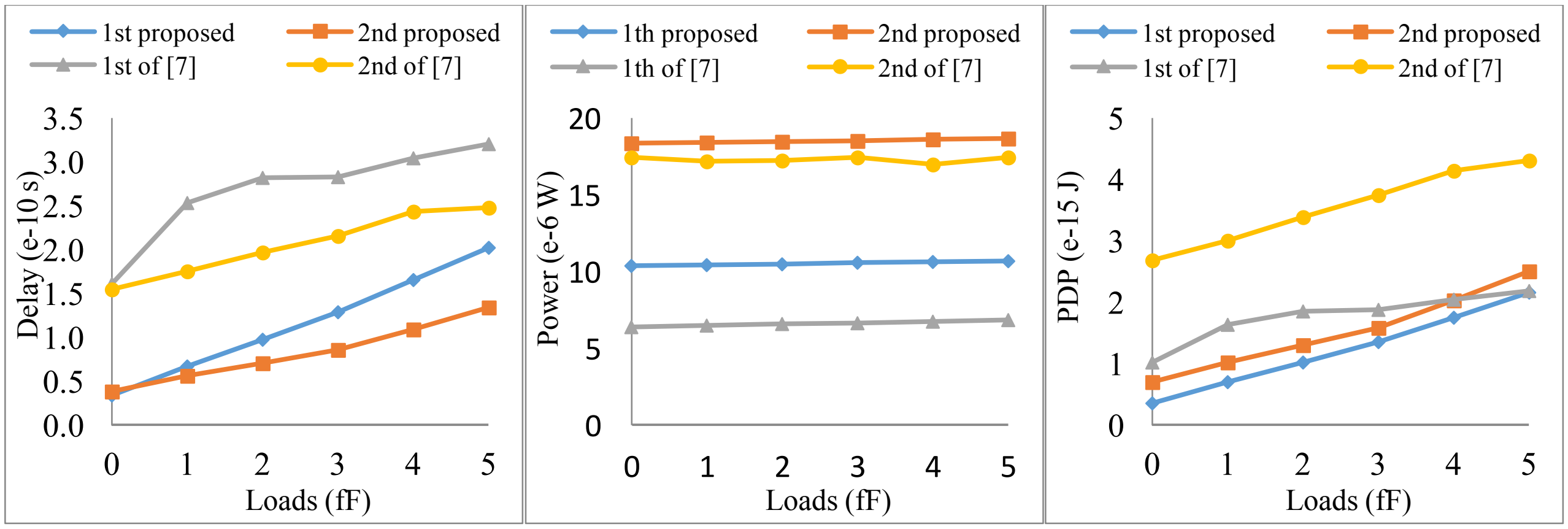

Figure 5: Delay, power and PDP of designs vs. load capacitors.

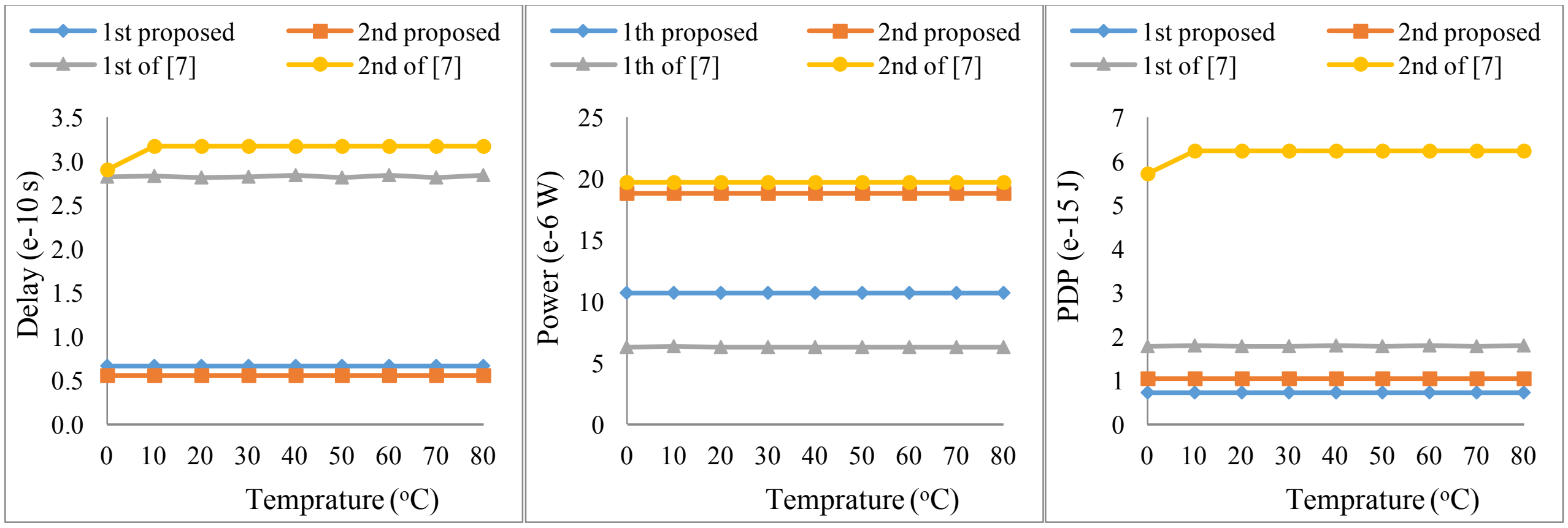

Figure 6: Delay, power and PDP of designs vs. temperature variation.

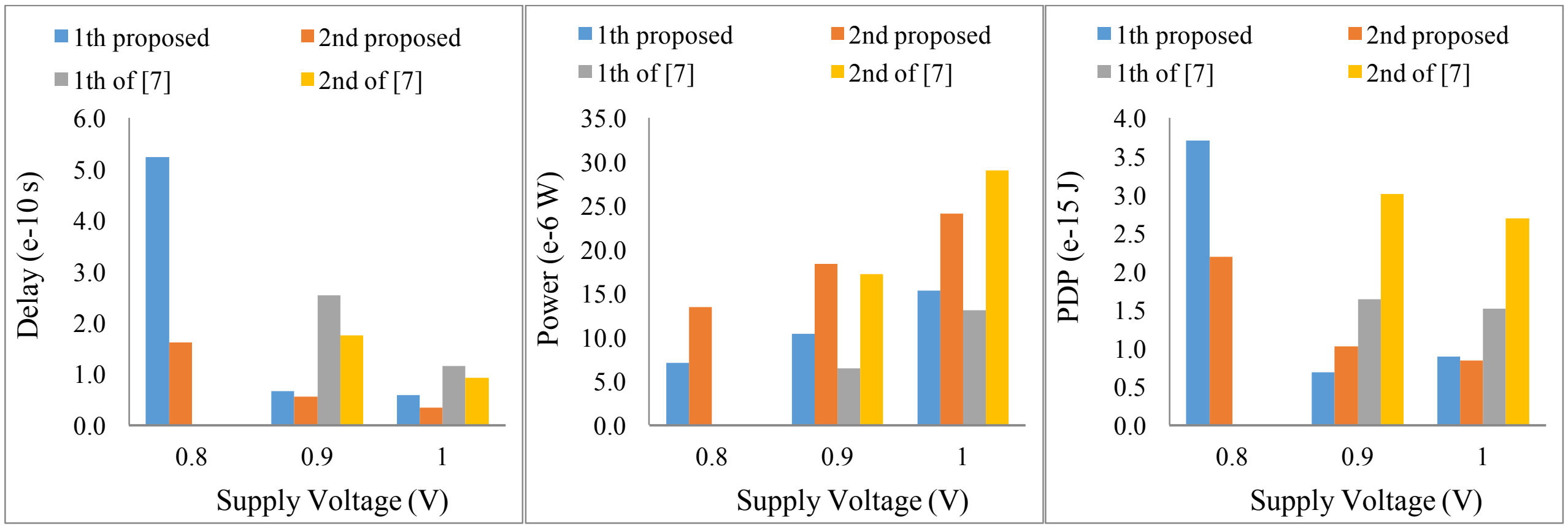

Figure 7: Delay, power and PDP of designs vs. supply voltage variation.